\documentclass[twocolumn,showpacs,secnumarabic,amssymb,nobibnotes,aps,prl]{revtex4-1}
\usepackage{amsmath}
\usepackage{latexsym}
\usepackage{float}
 \usepackage{amssymb}
\usepackage{graphicx}
\usepackage{textcomp}
\usepackage{hyperref}

 1
\def\ba{\begin{eqnarray}}
\def\ea{\end{eqnarray}}
\def\be{\begin{equation}}
\def\ee{\end{equation}}
\def\bm{\begin{math}}
\def\me{\end{math}}

\newcommand{\dummy}

\begin{document}

%\title{Clustering Dynamics in Non-sticky Granular Fluid}
%\title{ Strong Finite-size Effects during Coarsening in Non-sticky Granular fluid}
\title{Persistence in Ferromagnetic Ordering: Dependence upon initial configuration}
%\vskip 0.5cm
\author{ Saikat Chakraborty and Subir K. Das$^{*}$}
\affiliation{Theoretical Sciences Unit, Jawaharlal Nehru Centre for Advanced Scientific Research,
 Jakkur P.O, Bangalore 560064, India}

\date{\today}

\begin{abstract}
\par
We study the dynamics of ordering in ferromagnets via Monte
Carlo simulations of the Ising model, employing the Glauber spin-flip 
mechanism, in space dimensions $d=2$ and $3$. Results for the persistence probability and the domain
growth are discussed for quenches to various temperatures ($T_f$) 
below the critical one ($T_{c}$), from different
initial temperatures $T_{i} \geq T_{c}$. In long time limit, 
for $T_{i} > T_{c}$, the persistence probability exhibits power-law
decay with exponents $\theta \simeq 0.22$ and $\simeq 0.18$ in $d=2$ and $3$, respectively.
For finite $T_i$, the early time behavior is a different power-law whose
life-time diverges and exponent decreases as $T_{i} \rightarrow T_{c}$. 
The crossover length between the two steps diverges as the equilibrium correlation length. $T_i=T_c$ is 
expected to provide a {\it{new universality class}} for which we obtain $\theta \simeq 0.035$
in $d=2$ and $\simeq 0.10$ in $d=3$. 
The time dependence of the average domain size $\ell$, however, is observed to be rather
insensitive to the choice 
of $T_i$.

\end{abstract}

\maketitle
\section{\textrm{I} Introduction}
When a homogeneous system is quenched below the critical point,
the system becomes unstable to fluctuations and approaches
towards the new equilibrium via the formation and growth of particle
rich and particle poor domains \cite{onuki_phase,puri_phase,bray_phase,%
binder_phase}. In such nonequilibrium 
evolutions, over several decades, aspects that received significant
attention are the domain pattern \cite{bray_phase,cross,OJK,bray_puri,%
toyoki,das1}, rate of domain growth \cite{cross,LS,bind_stauf,allen_cahn,siggia,huse,suman1},
persistence \cite{satya_1,satya_2,derrida_1,derrida_2,manoj,jkb,saha} and 
aging \cite{d_fisher,liu,corberi_1,ahmad,suman_prl,midya}. Average size, $\ell$, of domains
typically grows with time ($t$) as \cite{cross} 
\begin{equation}\label{eq1}
 \ell \sim t^{\alpha}.
\end{equation}
The value of the exponent $\alpha$ for nonconserved order-parameter
dynamics \cite{cross,allen_cahn}, e.g., during ordering in an uniaxial
ferromagnet, is $1/2$, in space dimension $d=2$. In addition to the interesting structures exhibited
by the domains of like spins (or atomic magnets) in a 
ferromagnet, the unaffected or persistent spins also form beautiful
fractal patterns \cite{satya_1,satya_2,derrida_1,derrida_2,%
manoj,jkb}. Typically, fraction of such spins, 
henceforth will be referred to as the persistent probability, $P$, decays as
\begin{equation}\label{eq2}
 P \sim t^{-\theta},
\end{equation}
with \cite{manoj,stauffer} $\theta$ having a value $\simeq 0.22$ for the Ising model
in space dimension $d=2$ and $\simeq 0.18$ in $d=3$.
\par
The values of the exponents mentioned above are understood to be true for
the perfectly random initial configurations, mimicking the 
paramagnetic phase at temperature $T=\infty$. Another relevant situation
is to quench a system from finite initial temperature ($T_{i}$) 
with a large enough equilibrium correlation length $\xi$. However, this problem
has received only occasional  attention \cite{dasgupta,humayun,newman,blanchard}, though experimentally
very relevant. In this context, the behavior of the two-time equal point 
correlation function, relevant in the aging phenomena, was studied \cite{humayun,newman}
in $d=2$ for $T_i=T_c$, the critical temperature. It was pointed out 
that such quenches would form a {\it{new universality class}} and 
was shown that the decay of the above
correlation was significantly slower for $T_i=T_c$ than $T_i=\infty$. In view of that 
a slower decay of $P$ is also expected. Apriori, however, it is unclear whether there will be 
quantitative similarity between the degree of changes in the two quantities. 
On the other hand, the  behavior of $P$ and $\ell$ are expected to be disconnected \cite{leticia}. 
Nevertheless, the rate of growth of $\ell$ may be different for $T_i=T_c$ and $T_i= \infty$, 
at least during the transient period. 
In this paper, we address the $T_i$ dependence for persistence
and domain growth in a ferromagnet, via Monte Carlo (MC) simulations \cite{d_landau} 
of nearest neighbor Ising model \cite{d_landau}

\begin{equation}\label{eq3}
 H = -J\sum_{<ij>}S_i S_j;~S_{i}=\pm1,~J>0,
\end{equation}
in $d=2$ and $d=3$, on square and simple cubic lattices, respectively.
\par
Starting from a high value, as $T_{i}$ approaches 
$T_{c}$ [$\simeq$ $2.27 J/k_{B}$ in $d=2$ or $4.51 J/k_{B}$ in $d=3$, $k_{B}$ being the 
Boltzmann constant], a two-step decay in $P$ becomes prominent, except for
$T_{i}=T_{c}$. For the latter, power-law behavior
with exponents much smaller than the ones observed for quenches
from $T_i= \infty$ lives forever. In addition 
to identifying these facts, a primary 
objective of the paper is to accurately quantify these decays.
For the domain growth, on the other hand, we do not observe a modification to
time dependence with the variation of $T_i$, almost from the very beginning.
\par
The rest of the paper is organized as follows. In section \textrm{II} we briefly 
describe the methods. Results from both the dimensions are presented in section 
\textrm{III}. Section \textrm{IV} concludes the paper with a summary and outlook.
\section{\textrm{II} Methods}
The nonconserved order-parameter dynamics in the MC simulations
have been incorporated via the Glauber spin-flip mechanism \cite{glauber}.
In this method, a randomly chosen spin is tried for a change in
sign which is accepted according to the standard Metropolis algorithm \cite{d_landau}.
We apply periodic boundary
conditions in all directions. Time is expressed in units of MC steps (MCS),
each MCS consisting of $L^{d}$ trials, $L$ being the linear dimension
of a square box. We have computed $\ell$ from the domain size 
distribution, $P_{d}(\ell_{d},t)$, as \cite{suman1}
\begin{equation}\label{eq4}
 \ell(t)= \int{\ell_{d}P_{d}(l_{d},t)d\ell_{d}},
\end{equation}
where $\ell_{d}$ is calculated as the distance between two successive
interfaces in any direction. All lengths are expressed in units
of the lattice constant $a$. We present the results after averaging over
multiple initial configurations. This number ranges from $20$ (for $L=1024$) 
to $200$ (for $L=400$) in $d=2$ and $10$ (for $L=256$) to $50$ (for $L=64$) in $d=3$.
The initial configurations for $T_i$ close to $T_c$ 
were carefully prepared via very long runs. At $T_c$, depending upon the system size, 
length of such runs varied between $5 \times 10^6$ to $10^8$ MCS.
\section{\textrm{III} Results}
\subsection{A. $d=2$}
\begin{figure}[htb]
\centering
\includegraphics*[width=0.45\textwidth]{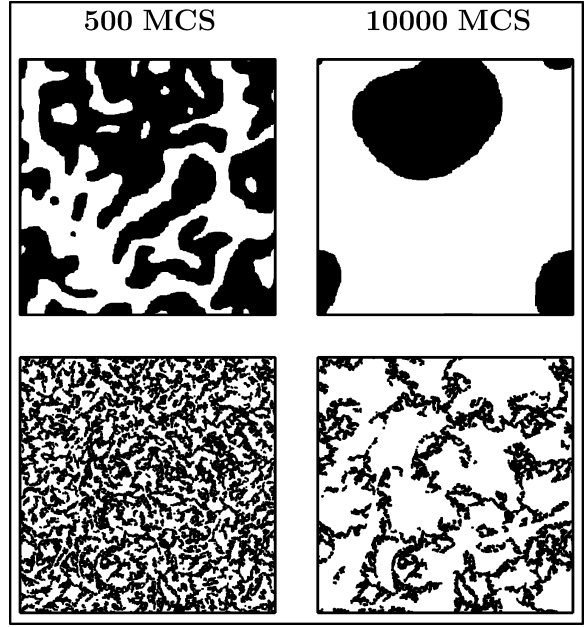}
\caption{\label{fig1} Upper panels show snapshots during the
evolution of the Glauber Ising model with $T_i=\infty$, $T_f=0$ and $L=512$.
The black regions represent domains of up spins. The lower panels show the unaffected spins,
marked in black, corresponding to the evolution snapshots above them. These results 
correspond to $d=2$.
}
%\vskip 0.75cm
%\includegraphics*[width=0.32\textwidth]{fig5b.eps}
%\caption{\label{fig5} (a) Snapshots showing only the unturned spin sites for quenches from random initial
%configuration to temperatures mentioned above
%each frame.All of these pictures are from $t = $ and $L = 512$.
%(b) Log-log plots of persistence probability, P, as a function of time, for quenches from random initial
%configuration to various temperatures, indicated
%on the figure. The solid line there has a power-law behavior $P \sim t^{-0.22}$.
%}
\end{figure}

\begin{figure}[htb]
 \centering
\includegraphics*[width=0.45\textwidth]{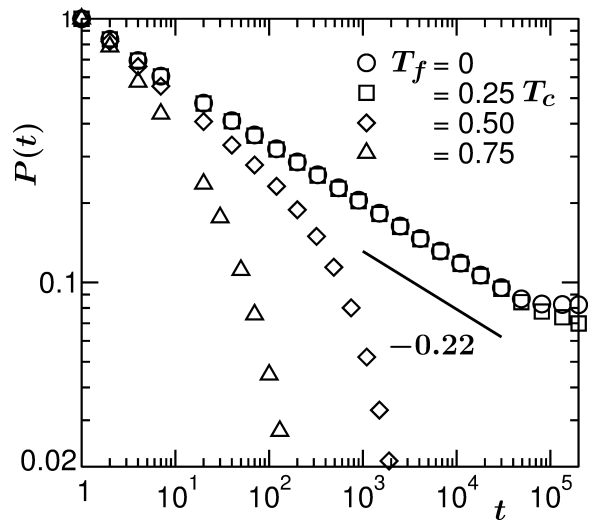}
\caption{\label{fig2} Plots of persistence probability, $P$, vs,
time, on a log-log scale, for quenches from $T_{i}= \infty$, with $L=512$, in $d=2$. 
Four different values of $T_{f}$ are included. The solid line
there has a power-law decay with exponent 0.22.}
\end{figure}

\begin{figure}[htb]
 \centering
\includegraphics*[width=0.45\textwidth]{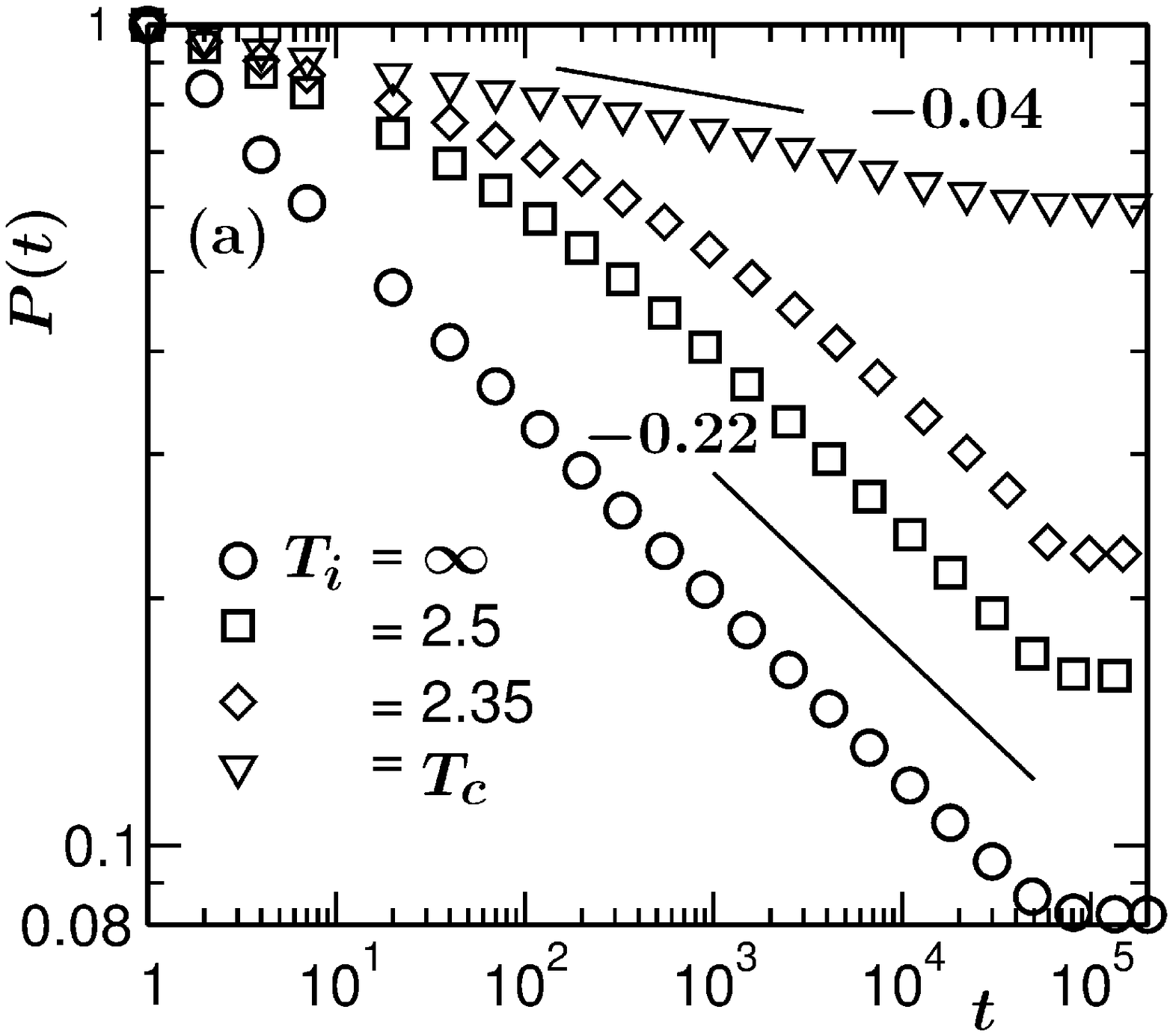}
\vskip 0.75cm
\includegraphics*[width=0.45\textwidth]{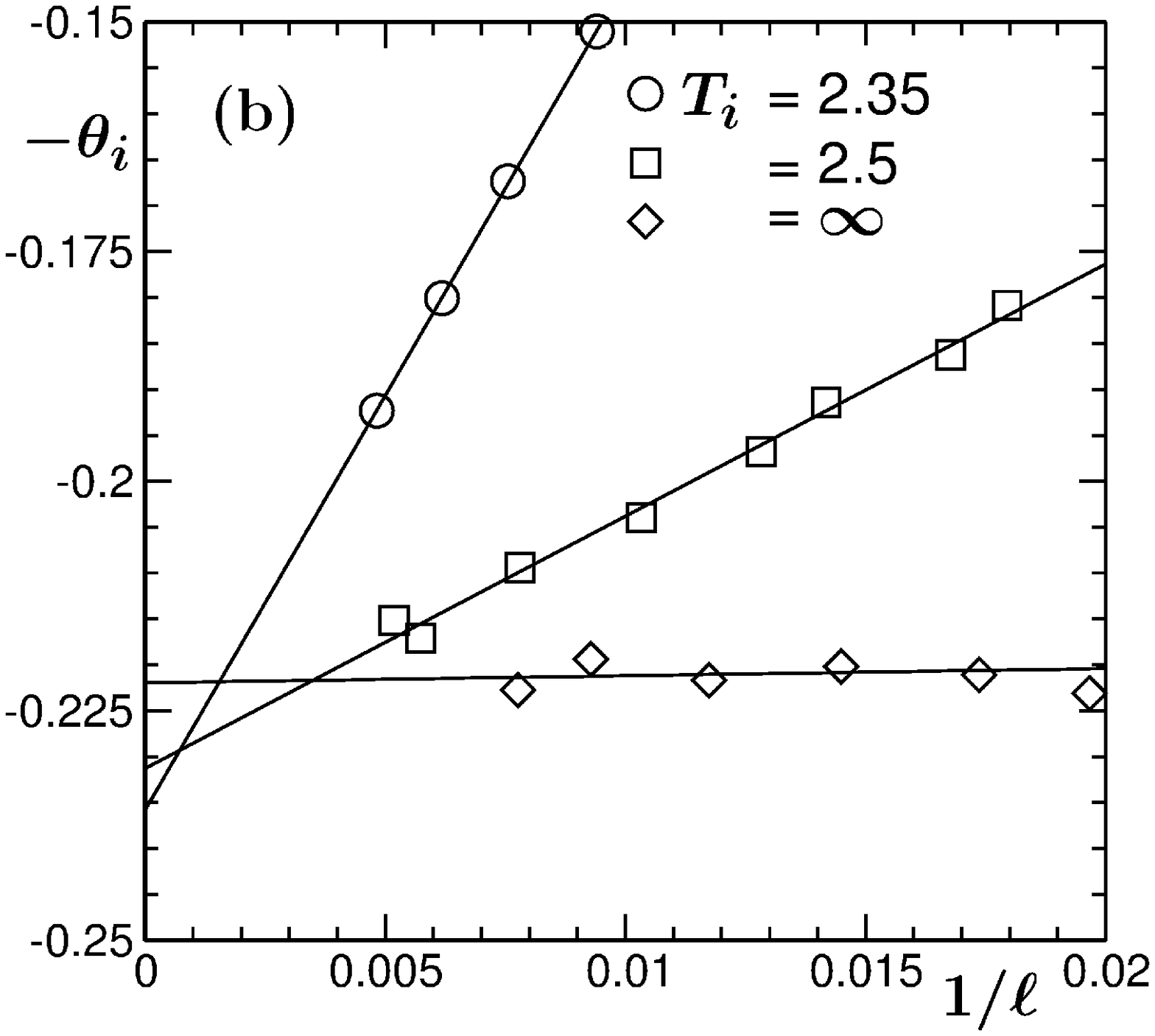}
\caption{\label{fig3} Log-log plots of $P$ vs $t$, for quenches
from different values of $T_{i} (\geqslant T_{c})$, to $T_{f}=0$. 
Continuous lines there correspond to power-law decays with exponents $0.22$ and $0.04$.
(b) Instantaneous exponents $\theta_{i}$ are plotted vs $1/\ell$, for
the quenches in (a), excluding $T_i=T_c$ case. Here we have
included only the late time behavior. The continuous lines in this figure are linear 
fits, providing the values of $\theta$ as the ordinate intercepts. All 
results are from simulations in $d=2$.
}
\end{figure}

\begin{figure}[htb]
 \centering
\includegraphics*[width=0.45\textwidth]{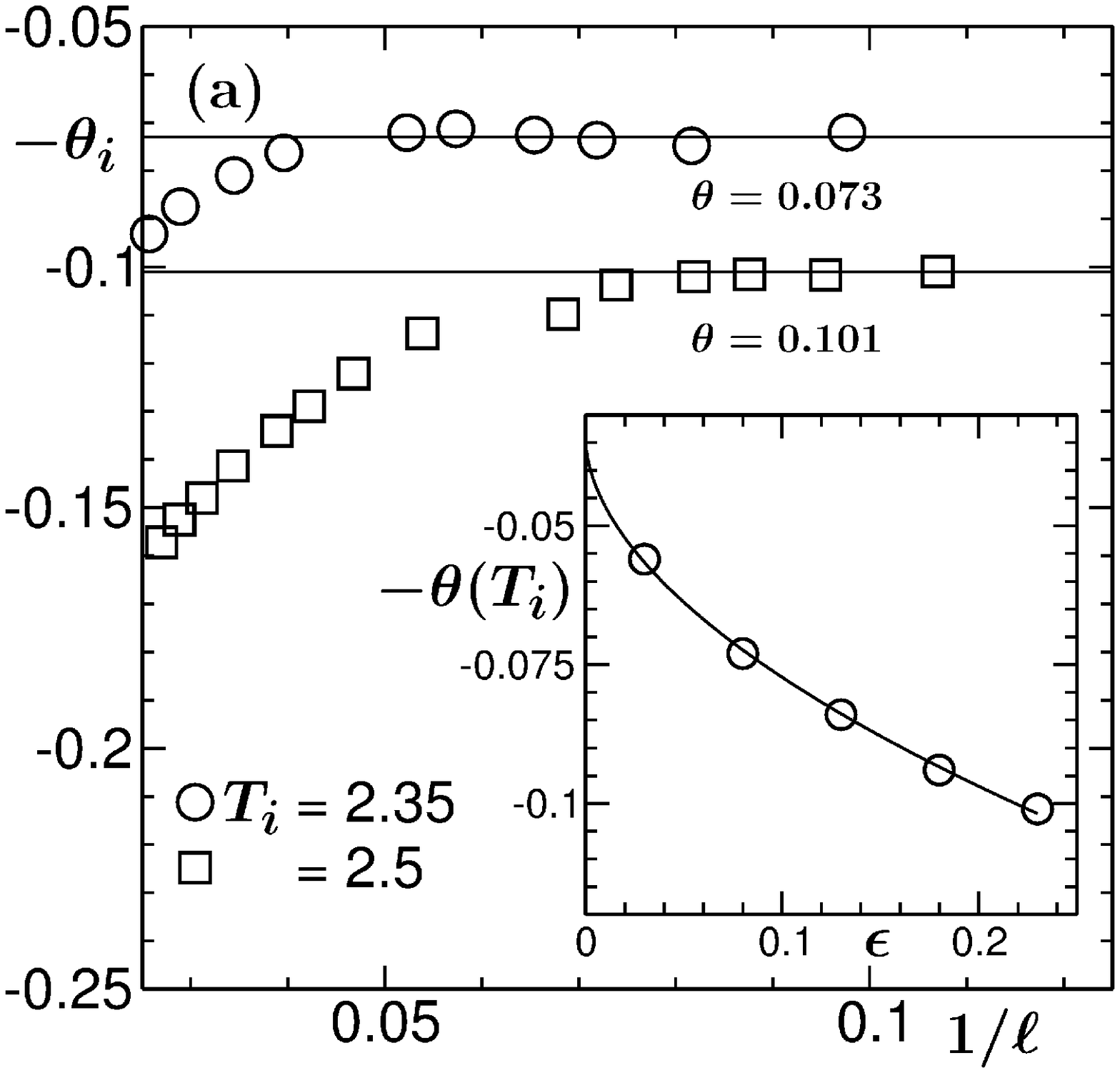}
\vskip 0.75cm
\includegraphics*[width=0.45\textwidth]{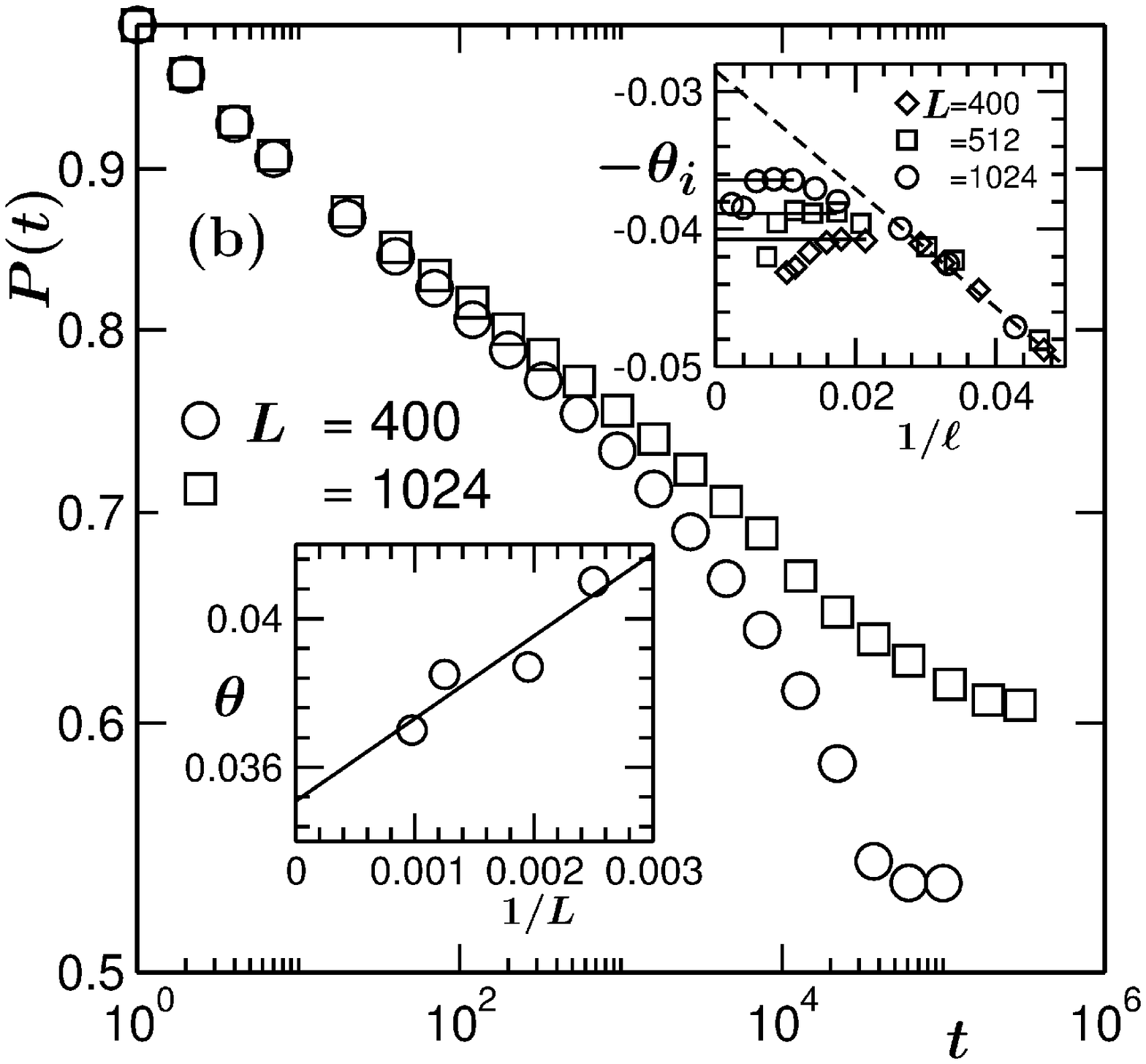}
\caption{\label{fig4} (a) Instantaneous exponents $\theta_{i}$ are plotted vs $1/\ell$
for two of the quenches in Fig. \ref{fig3}(a). Here we have focussed
on the first step of the decays, exponents for which are obtained from
the flat regions, marked by the horizontal solid lines. 
These values of $\theta$ are plotted, vs $\epsilon=T_i-T_c$, in the inset of the figure. 
The continuous line there is a power-law fit (see text).
(b) Plots of $P$ vs $t$, for two different system sizes, with $T_i=T_c$ and $T_f=0$. 
The upper inset shows $\theta_i$ vs $1/\ell$ for three different values of $L$. 
The dashed line in this inset is a linear extrapolation using data in the small $\ell$ region.  
The $L$-dependent exponents obtained from flat regions of the plots (see the horizontal 
solid lines) are plotted vs $1/L$ in the lower inset. The solid line there is a linear fit.
All results correspond to $d=2$.
}
%for early time power-law decay, }
%$P(t)$ is plotted vs $t$, for quench from $T_{i}=T_{c}$ to $T_{f}=0$, on log-log scales. The solid line has a 
%power-law exponent $-0.04$.}
\end{figure}

%\begin{figure}[htb]
% \centering
%\includegraphics*[width=0.45\textwidth]{fig5a.eps}
%\vskip 0.75cm
%\includegraphics*[width=0.45\textwidth]{fig5b.eps}
%\caption{\label{fig5} (a) Plot of $P(t)$ vs $t$, on log-log scales, for a quench from $T_{i}= \infty$ to $T_{f}=0.5T_{c}$. Here a modified
%definition is used for the calculation of $P$. Essentially, spins affected by thermal fluctuations are discarded from the calculation. 
%Inset shows $\theta_{i}$ as a function of $1/\ell$, for the quench in the main frame. The solid line there is a linear extrapolation to
%$\ell = \infty$.
%(b) Same as (a) but for $T_{i}=T_{c}$ and $T_{f}=0.5T_{c}$. In the main frame here we have included results with the modified definition.}
%\end{figure}

%\begin{figure}[htb]
% \centering
%\includegraphics*[width=0.45\textwidth]{fig6.eps}
%\caption{\label{fig6} Plots of instantaneous exponent, $\alpha_{i}$, vs $1/\ell$, for $T_{i} \infty $. Three values of $T_{f}$ are 
%chosen. The solid horizontal line represents $\alpha=1/2$.}
%\end{figure}

\par
Growth of the domains have been demonstrated in the upper frames of Fig. \ref{fig1} 
for the system size $L=512$ in $d=2$. 
In this figure we show snapshots from two different times during the evolution
of the Glauber Ising model. In the lower frames of the figure, 
we show pictures marking only the persistent spins.  Beautiful patterns are visible. 
These results correspond to a quench 
from $T_{i}= \infty$ to $T_{f}=0$.
\par
Plots of $P$, for $T_{i}= \infty$ and few different values  
of $T_{f}$ are shown vs t in Fig. \ref{fig2}. The data for
$T_f=0$ and $0.25T_{c}$ are consistent with each other and follow
power-law, the exponent being $\theta \simeq 0.22$. 
The flat behavior at the end is due to the finite-size effects. 
This value of $\theta$ is consistent with the observation by Manoj and Ray \cite{manoj}. However,
for higher values of $T_{f}$, as also previously observed \cite{derrida_1,derrida_2}, the decay is
not of power-law type. This is thought to be due to thermal fluctuation. 
When this fluctuation is taken care of, we observe $\theta \simeq 0.22$ 
for all the values of $T_f$ included in Fig. \ref{fig2}, in agreement with Ref. [21].
It is thought that persistence and domain growth are not strongly connected 
to each other. Interestingly, different behavior in Fig. \ref{fig2} for $T > 0.25T_c$ is 
strongly reflected in the domain growth also. For brevity we do not present it here.
\par
In Fig. \ref{fig3}(a) we show $P$ vs t plots, on a log-log scale,
for quenches to $T_{f}=0$, from a few different values of $T_{i}$,
all for the same system size $L=512$.
It appears that, in the long time limit, for $T_i > T_c$,
the decay is power-law, with the same exponent 
$\theta \simeq 0.22$. However, crossover to this functional form
gets delayed as $T_{i}$ approaches $T_{c}$. In fact, in the pre-crossover 
regime, another power-law decay, with smaller value of $\theta$, 
becomes prominent with the decrease of $T_{i}$. Such
a slower decay becomes ever-lived for $T_{i}=T_{c}$.
\par
In Fig. \ref{fig3}(b), we present the instantaneous exponent,
$\theta_{i}$, calculated as \cite{huse,suman1}
\begin{equation}\label{eq5}
 \theta_{i}=-\frac{d \ln P}{d\ln t},
\end{equation}
vs $1/\ell$, with the objective of accurate quantification of the 
second step of the decays for $T_i$ close to, but greater than, $T_c$. 
For the abscissa variable we have adopted $1/\ell$,  
instead of $1/t$, to visualize the long time limit better.
Data sets for each $T_i$ are fitted to linear functions to obtain the exponents. 
Within statistical error, for all the presented temperatures, it appears 
that $\theta$ is consistent with the quench from $T_i=\infty$ to $T_f=0$. 
Nonzero slopes in these plots are due to the presence of off-sets at the 
crossover times.
\par
Next we move to identify the exponent for $T_{i}=T_{c}$ and $T_f=0$. 
In Fig. \ref{fig3}(a), it appears that the $T_i=T_c$ data are reasonably consistent 
with $\theta=0.04$. Nevertheless, before the final finite-size effects appear (showing flat 
nature at very late time), there is a faster decay, albeit for a brief period. 
This can well be due to the fact that for a finite system, $\xi$ 
is not infinite at $T=T_c$, effectively implying that 
the initial configurations are prepared away from $T_c$. 
Thus, in this problem finite-size effects have two sources. 
One coming from the finiteness of the equilibrium correlation length, 
other being faced when the nonequilibrium domain size is close to the system size.
Thus, a quantification of the exponent $\theta$, for $T_i=T_c$, 
via finite-size scaling \cite{m_fisher}, becomes 
a challenging task. 
However, we appropriately take care of the shortcoming below, in 
various different ways which provide results consistent with each other. 
\par
In Fig. \ref{fig4}(a) we show the instantaneous exponents $\theta_i$, vs $1/\ell$, 
with the objective of quantifying 
the first step of the decay, for two values of $T_i$, close enough to $T_c$. 
As demonstrated, from the flat regions we identify the exponents 
for the early time or the first step of the decay for various values 
of $T_i$. These numbers are plotted in the inset of this figure 
as a function of $\epsilon= T_{i}-T_{c}$. The continuous line there is a fit
to the form 
\begin{equation}\label{eq6}
 \theta(T_{i})= \theta(T_{c})+A \epsilon^{x},
\end{equation}
providing $\theta(T_{i}=T_{c})=0.034$, $A=0.15$ and $x=0.54$.
Recall that this value of $\theta$ is the only decay exponent for $T_i=T_c$.
\par
To verify the above value of $\theta$ further, in Fig. \ref{fig4}(b) we 
present an exercise with different system sizes. In 
the main frame of this figure, we present $P$ vs $t$ data, 
for $T_{i}=T_{c}$, from two different values of $L$. It is seen 
that with the increase of the system size, there is a tendency of the data to settle down 
to a power-law for a longer period of time, following a marginally
faster decay at early time. In the upper inset of this figure we 
show $\theta_i$ vs $1/\ell$ for three different system sizes with $T_{i}=T_{c}$. 
The early time behavior appears linear, extrapolation of 
which leads to $\theta \simeq 0.029$. However, if 
the data in the main frame is closely examined, as already mentioned above, this part corresponds to the 
preasymptotic behavior, thus, should be discarded from the analysis. Actual asymptotic 
exponents should be extracted from the flat regions of the plots. 
The numbers obtained from these flat parts, 
as discussed, differs due to the finite-size effects
and thus, should be extrapolated to $L=\infty$ appropriately. 
These values are plotted in the lower inset as a function of $1/L$. A very reasonable linear fit 
(see the solid line) is obtained, providing $\theta = 0.035$. On the other hand, a 
nonlinear fit appears to be nearly quadratic providing $\theta = 0.037$. 
From all these exercises we conclude that $\theta (T_{i}=T_{c})=0.035 \pm 0.005$. 
This picture remains true for quenches from $T_c$ to nonzero values 
of $T_f$, if thermal fluctuation effects are appropriately taken care of. 
On this issue of thermal fluctuation, here, as well as for $T_i= \infty$, our studies were
restricted to only small system sizes. 
\par
The decay of the two-time correlation is also 
of power-law type. For quenches from $T_i=T_c$, the value of the exponent 
gets reduced by a factor $\simeq 10$, compared to $T_i= \infty$. In the present  
case the reduction factor is $\simeq 6.3$. While there may be connection 
between the two phenomena, but a search for matching between the two factors may 
not be justified. This is because, only spins with odd number of flips are 
important for the decay of the two-time correlation function.

\par
\begin{figure}[htb]
 \centering
\includegraphics*[width=0.45\textwidth]{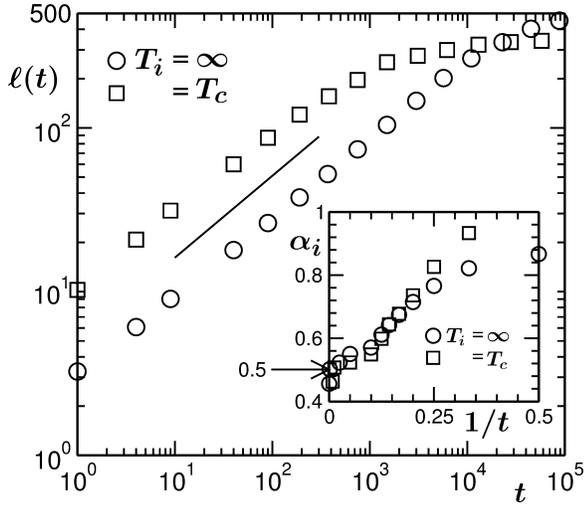}
\caption{\label{fig5} Average domain sizes, $\ell(t)$, are plotted vs $t$, for
quenches to $T_f=0$ from $T_i=\infty$ and $T_c$, in $d=2$. The solid line represents $t^{1/2}$ 
behavior. Corresponding instantaneous exponents, vs $1/t$, are shown in the inset. 
All these results are for $L=512$.
}
\end{figure}

It is certainly important to ask, if, like the decay of the persistence
probability and the two-time correlation \cite{humayun}, the growth of the average domain size 
also exhibits initial temperature dependence, at least at the transient level. Here we only
examine the cases $T_{i}= \infty$ and $T_{i}=T_{c}$, for quenches to $T_f=0$. 
In Fig. \ref{fig5} we present the $\ell$ vs $t$ plots for these two cases using a log-log scale. 
Both the data sets appear to grow slower than $t^{1/2}$. This can well be 
due to the presence of significantly big initial length 
$\ell_0$, which we examine below. While from this
figure it is difficult to identify any difference in the growth exponent  
between the two cases, there certainly exists difference in the finite-size effects, 
noting that $L=512$ in both the cases. 
\par
To learn better about the exponents, in the inset we present the instantaneous 
exponents \cite{huse,suman1} 
\begin{equation}\label{eq7}
 \alpha_{i}=\frac{d \ln \ell}{d\ln t},
\end{equation}
with the variation of t. Here, while calculating $\alpha_i$, we have
subtracted $\ell_0$ which are $\simeq 2$ 
and $\simeq 6.65$, respectively, for $T_{i}= \infty$ and $T_c$. 
This subtraction is meaningful, considering the fact that the pure scaling with 
respect to time is contained in $\ell-\ell_0$. Calculation of $\alpha_i$, 
without such subtraction, provided early time exponents much smaller than 
the theoretical expectation for the conserved dynamics \cite{huse}. 
This has previously been understood 
to be due to the curvature dependent correction. 
Such confusion has recently been corrected \cite{suman1}. 
Note that, there may be a delay time for a system to become unstable following 
a quench. Thus, for an appropriate understanding of a 
time dependent exponent, the value of $\ell_0$ 
need not be the length at $t=0$. Via finite-size scaling analysis, 
this was demonstrated in a recent work \cite{suman1}. 
Here, however, we do not undertake such a task.  
\par
It may be argued that 
the value of $\ell_0$ should be of the order of system size for $T_i=T_c$,
since $\xi \sim L$ at $T_c$. Note here that at criticality fluctuations 
exist at all length scales. At $T_i=T_c$, small value of $\ell_0$, compared to
$L$, is due to the fact that our calculations did not ignore such fluctuations.
%\begin{equation}\label{eq8}
% S(q \rightarrow 0,t)=\frac{k_{B}T\chi}{1+q^{2}\xi^{2}},
%\end{equation}
\par
First important observation from the inset of Fig. \ref{fig5} 
is that the value of $\alpha_i$ approaches $1/2$ 
from the upper side. This is in contradiction with the corresponding behavior for
the conserved order parameter dynamics with $T_f$ very close to 
zero \cite{suman2}. In the latter case, the early time 
dynamics provides a growth exponent much
smaller than the expected asymptotic value $1/3$. 
Second, after $t=5$, both the data sets practically 
follow each other, implying no difference in the growth of $\ell$
almost from the beginning!

\par
From the length (or time) dependence of $\alpha_i$, one can write 
\begin{equation}\label{eq8}
 \alpha_i=\alpha + f(1/\ell),
\end{equation}
to obtain
\begin{equation}\label{eq9}
 \int{\frac{d\ell}{{\alpha}l\big[1+\frac{1}{\alpha}f(1/\ell)\big]}} = \ln t.
\end{equation}
If $f(1/\ell)$ can be quantified accurately from the simulation data, 
a full time dependence of $\ell$ is obtainable. 
E.g., if $f(1/\ell)$ is a power law, $ A_{\beta}/\ell^{\beta}$, $A_{\beta}$
being a constant, by taking ${\alpha}l^{\beta} > A_{\beta}$, one finds
\begin{equation}\label{eq10}
 \ln \frac{\ell^{1/\alpha}}{t} \sim \frac{1}{{\alpha^2}{\beta}{\ell^{\beta}}}.
\end{equation}
Assuming that a correction disappears fast, such that $\ell \simeq t^{\alpha}$, 
we obtain 
\begin{equation}\label{eq11}
 \ell \sim t^{\alpha}\exp(-\frac{C}{{\alpha}{\beta}t^{{\alpha}{\beta}}}),
\end{equation}
C being a constant.
Such full forms are useful for a finite-size scaling analysis to accurately
quantify the exponent $\alpha$. It appears 
that even for a power-law behavior of $f(1/\ell)$, the asymptotic behavior 
in the growth law is reached exponentially fast. Of course, from least square
fitting exercise of the $\ell$ vs $t$ data also one can aim to obtain the early time
corrections. However, this method is more arbitrary. Often derivatives help
guessing the functional forms better.
We leave the exercise of a finite-size scaling analysis to accurately estimate $\alpha$ 
and the finite-size effects for a future work.
\par
Before moving on to presenting results in $d=3$, we discuss the issue of persistence again.
The essential feature in the initial configurations prepared at different temperatures
is the difference in the equilibrium correlation length $\xi$. The basic question, prior to the study, one asks,
how the size of $\xi$ affects the decay of persistence probability? For each value of $\xi$, do
we have a unique exponent $\theta$ describing the full decay? The answer, as we have observed, 
is certainly not in affirmative. Essentially, the decay exponent for $T_i=\infty$ is recovered for all 
$\xi$ $(<\infty)$ in the long time limit. The crossover to this asymptotic behavior 
gets delayed with the increase of $\xi$. It is then natural to ask if this crossover 
occurs when $\ell$ crosses $\xi$.
This can be answered by appropriately estimating the crossover length $\ell_c$.

\begin{figure}[htb]
 \centering
\includegraphics*[width=0.45\textwidth]{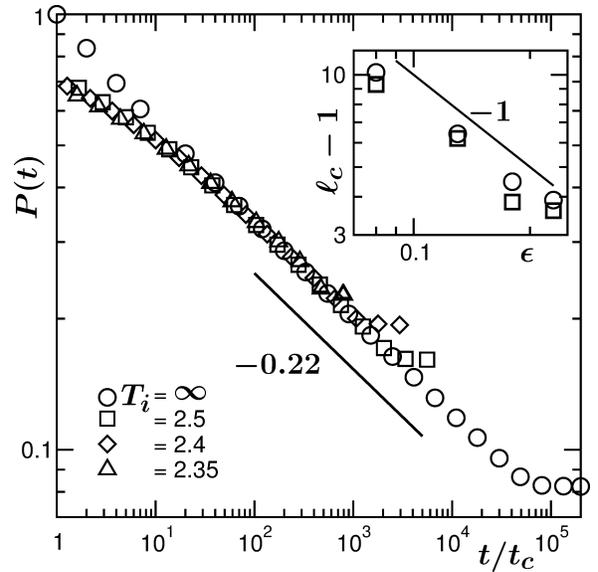}
\caption{\label{fig6} Scaling plot of persistence probability versus $t/t_c$ where the 
crossover time (to the asymptotic behavior) $t_c$ has been used as an adjustable 
parameter to obtain optimum data collapse.
Inset: Plots of $\ell_{c}-1$ versus $\epsilon$. The circles correspond to estimates
of $\ell_c$ from $t_c$, the squares are directly obtained from the scaling plots of 
$P$ vs $\ell/\ell_c$. The solid line has $d=2$ Ising critical divergence of correlation length.
All results were obtained using $L=512$
in $d=2$.
}
\end{figure}

\par
In the main frame of Fig. \ref{fig6} we show plots of persistence from different 
values of $T_i$, for quenches to $T_f=0$. Here the time axis is scaled by
appropriate factors (expected to be proportional to cross over time $t_c$)
to obtain collapse of data in the asymptotic regime. Quality of collapse, on top of the
$T_i= \infty$ data set, again confirms that $\theta \simeq 0.22$ in the $t \rightarrow \infty$
limit for all $T_i$ ($ > T_c$). From the square roots of these $T_i$ dependent scaling factors,
one can obtain $\ell_c$ which is expected to scale as $\ell_c \sim \xi \sim \epsilon^{-\nu}$. 
Note that for the Ising model $\nu=1$ in $d=2$ and $\simeq 0.63$ in $d=3$. 
considering that the $T_i=\infty$
data have been used as the reference case, it will be appropriate to fit the data set for
$\ell_c$ to the form
\begin{equation}\label{eq12}
 \ell_c= 1+A_c\epsilon^{-\nu},
\end{equation}
since $\ell_c \rightarrow 1$ for $T_i=\infty$. Unless we are very close to $T_c$ such additional
term cannot be neglected. In the inset of Fig. \ref{fig6}
we ahve plotted $\ell_{c}-1$ as a function of $\epsilon$, on a log-log scale. The data set (circles)
appear consistent with $\nu=1$. When $\ell_c$ is extracted from $t_c$, a 
better exercise requires incorporation of $\ell_0$ and growth amplitude for each $T_i$.
To avoid this problem, we have also obtained $\ell_c$ directly from the scaling 
plots of persistence data vs $\ell/\ell_c$ (see the squares). Both data sets 
appear nicely consistent with each other. In fact, fitting these data to the form 
in Eq. (\ref{eq12}) we obtain $\nu \simeq 0.95$.
\subsection{B. $d=3$}

\begin{figure}[htb]
 \centering
\includegraphics*[width=0.45\textwidth]{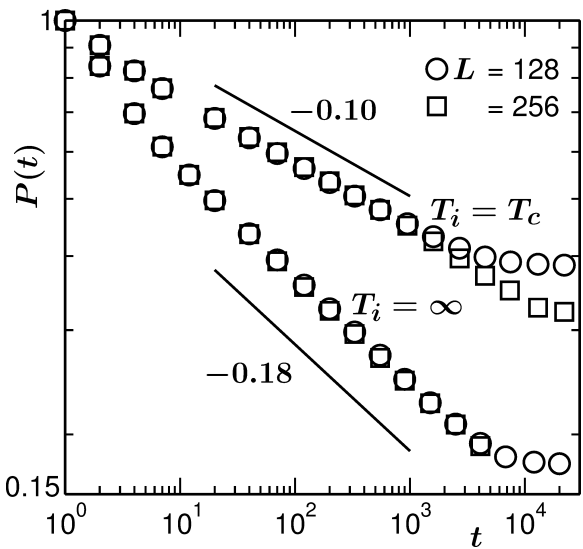}
\caption{\label{fig7} Plot of $P$ vs $t$, for quenches from $T_i=\infty$ and $T_i=T_c$,
 to $T_f=0$. In each of the cases results from two different system sizes are included.
 The solid lines have power-law decays with exponents 0.1 and 0.18, as indicated on the 
 figure. All results corresponds to $d=3$.
}
\end{figure}

\begin{figure}[htb]
 \centering
\includegraphics*[width=0.45\textwidth]{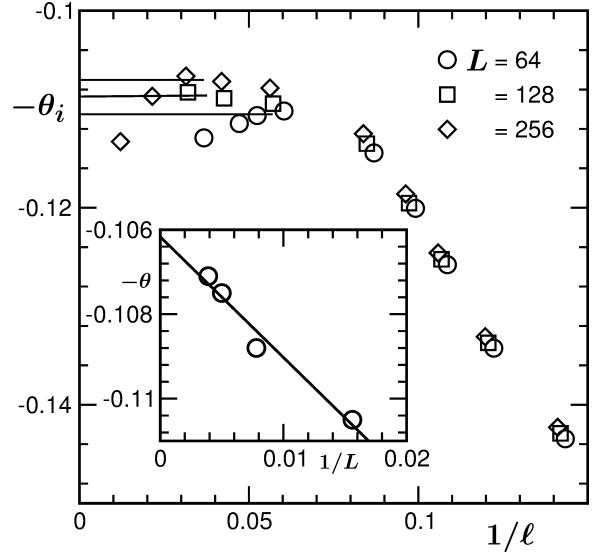}
\caption{\label{fig8} Instantaneous exponents $\theta_i$ are plotted vs $1/\ell$, for
quenches from $T_i=T_c$ to $T_f=0$ in $d=3$. Results from different values of L are included.
The horizontal solid lines are related to the estimation of L-dependent $\theta$.
Inset: System size dependent $\theta$ are plotted 
as a function of $1/L$. The continuous line there is a linear fitting (see text for details).
}
\end{figure}

\begin{figure}[htb]
 \centering
\includegraphics*[width=0.45\textwidth]{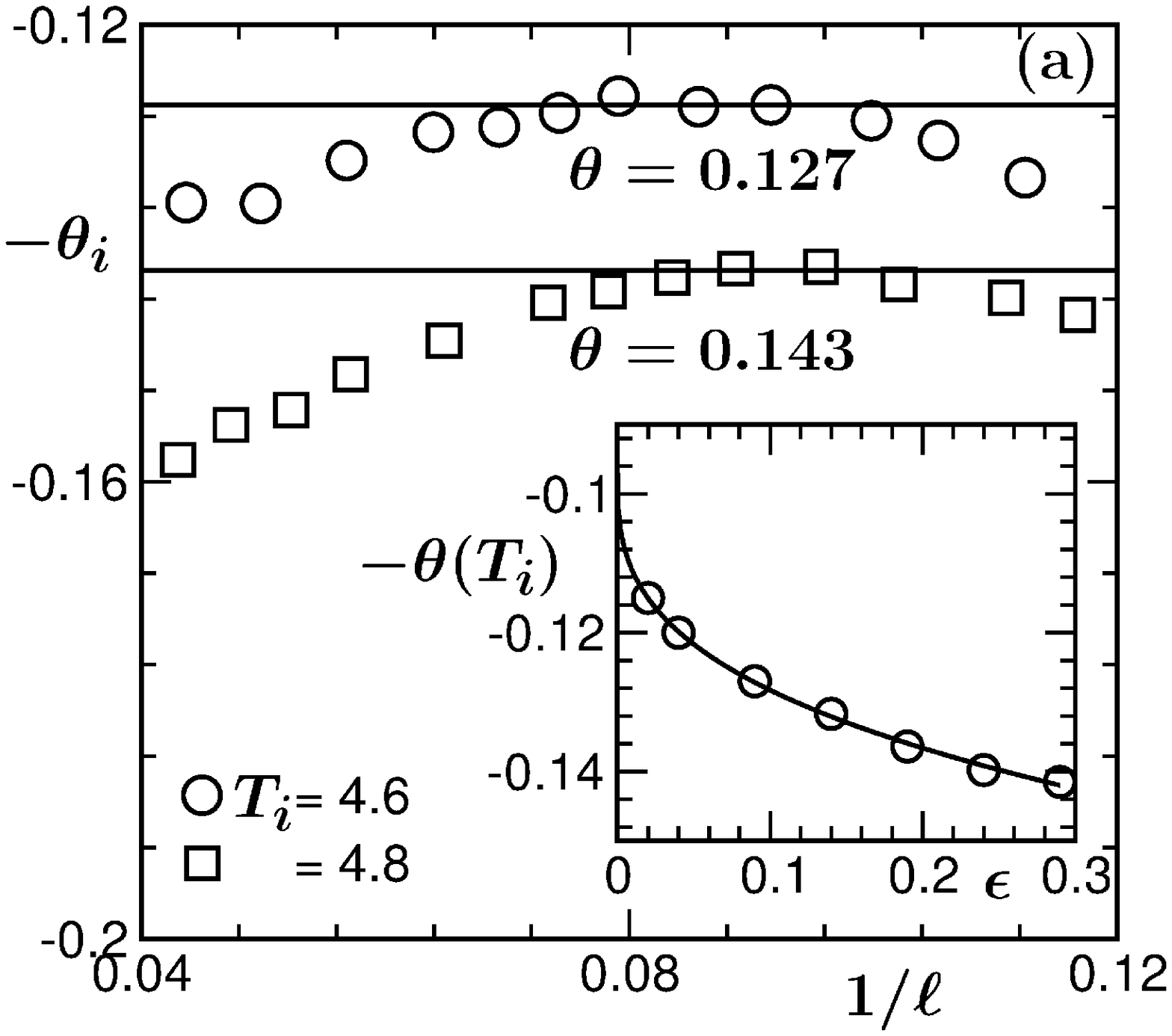}
\vskip 0.75cm
\includegraphics*[width=0.45\textwidth]{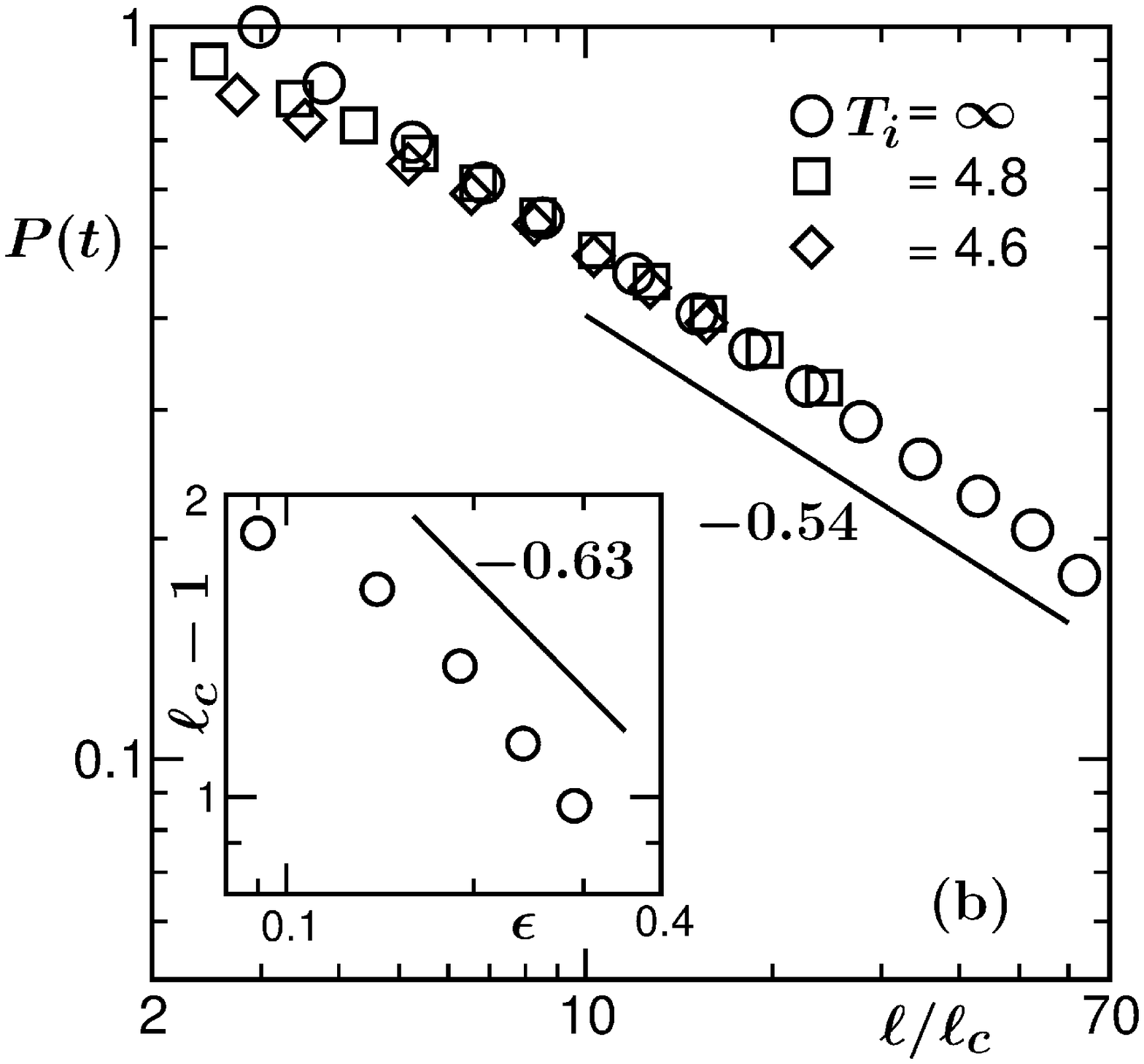}

\caption{\label{fig9}(a) Estimation of $\theta$ for the first step of decay is demonstrated in $d=3$.
Inset: Exponent $\theta$ for the first step of the decay is plotted as a function of
$\epsilon$, in $d=3$. The continuous line is a non-linear fitting. Further details are provided in the text.
(b) Scaling plot of $P$, versus $\ell/\ell_c$, in $d=3$. The solid line has a power-law decay with exponent $0.54$.
Inset: Plot of $\ell_{c}-1$, in $d=3$, versus $\epsilon$. The solid line there has $d=3$ Ising 
critical divergence. All results were obtained using $L=128$.
}
\end{figure}

\par
Next we present results in $d=3$. All the important facts being discussed in the
previous subsection, here we straightway present the results.
Noting that nothing remarkable happened for domain 
growth in the lower dimension, we stick only to the persistence probability.
\par
In Fig. \ref{fig7} we show the $P$ vs $t$ plots for quenches from $T_i= \infty$ and $T_i =T_c$,
keeping $T_f=0$ in both the cases. For each value of $T_i$, results from multiple system sizes 
are presented. The data for $T_i= \infty$ are consistent with $\theta=0.18$, previously
reported in \cite{stauffer}. Thus, here we aim to accurately quantify the value of $\theta$
for $T_i=T_c$.
\par
Even though, for $T_i=T_c$, data from both the system sizes in Fig. \ref{fig7} look consistent
with each other, finite-size effects are detectable from a closer look. In the main frame 
of Fig. \ref{fig8} we plot $\theta_i$ versus $1/\ell$ for a few different values of $L$.
Like in $d=2$, from the 
flat regions we identify system size dependent $\theta$, a plot of which is shown in the 
inset of this figure. Again, the $\theta$ vs $1/L$ data exhibits a reasonable linear trend and 
an extrapolation to $L= \infty$ provides $\theta \simeq 0.106$.
\par
Similar to $d=2$, for $T_c<T_i< \infty$, two step decays exist in $d=3$ as well.
In the main frame of Fig. \ref{fig9}(a) we have demonstrated the estimation of $\theta$
for the first step, for two representative values of $T_i$.
In the inset of Fig. \ref{fig9}(a) we have plotted these exponents 
as a function of $\epsilon$. A fit of this data set to the form in Eq. (\ref{eq6})
provides $\theta (T_i=T_c)=0.096$, $A=0.07$ and $x=0.34$. This value of $\theta$ is in good
agreement with the one obtained from Fig. \ref{fig8}. Thus, in $d=3$, for $T_i=T_c$,
we quote $\theta=0.10 \pm 0.02$. The effect of growing correlation length in the initial
configurations certainly appears weaker in this space dimension.
\par
In Fig. \ref{fig9}(b) we show scaling plots of $P$, vs $\ell/\ell_c$, using data from different values of $T_i$.
Collapse of data is good. The late time behavior is power-law and is consistent with a
decay exponent $0.54$. Considering that $\theta \simeq 0.18$ in $d=3$, this implies $\alpha=1/3$
in $d=3$ \cite{sire}. As stated in Ref. [39], deviation of $\alpha$, in this dimension, from $1/2$,
is not yet understood. To avoid this, as well as to get rid of the influence of $T_i$ dependent $\ell_0$
and growth amplitude, we have obtained $\ell_c$ from these plots only. 
\par
Finally in the inset of Fig. \ref{fig9}(b) we plot $\ell_{c}-1$
as a function of $\epsilon$, on a log-log scale.
The divergence of the length scale is consistent with a power law exponent $0.63$
which is the critical exponent for $\xi$ in $d=3$. In fact, a fitting to the form in 
Eq. (\ref{eq12}), excluding the point closest to $T_c$, provides $\nu \simeq 0.67$. 
The reason for the exclusion of the last point is finite-size effect which is clearly visible.
Data below $\epsilon=0.1$ are, however, included in the nonlinear fitting in the 
inset of Fig. \ref{fig9}(a). This is the reason for quoting larger error bar in the
final value of $\theta$ there.
\section{\textrm{IV} Conclusions}
\par
In conclusion, we have studied phase ordering dynamics in
Ising ferromagnets for various combinations of
initial ($T_i$) and final ($T_f$) temperatures in $d=2$ and $3$. In this work,
the primary focus has been on the persistence probability, $P$, and its connection 
with the growth of average domain size, $\ell$, as well as the equilibrium 
initial correlation length $\xi$.
\par
Our general observation has been that, irrespective of the value
of $T_i$, the decay of $P$ becomes faster with the increase of $T_f$, 
after a certain critical number for the latter. 
This is understood to be due to spins affected by thermal fluctuations.
When this effect is taken care of \cite{derrida_2}, the long
time decay appears to be power law with exponent \cite{manoj,stauffer} consistent 
with the one for quench to $T_f=0$.
\par
As $T_i$ approaches $T_c$, two-step power-law decay becomes prominent,
the second part having exponent $\theta\simeq 0.22$ in $d=2$ and $\simeq 0.18$
in $d=3$, same as $T_i=\infty$ and $T_f=0$ case. For $T_i=T_c$, thought to 
provide a new universality class, the first part of the
two-step process lives for ever. The corresponding values of the exponent 
have been identified to be $\simeq 0.035$ in $d=2$ and $\simeq 0.10$ in
$d=3$. Thus, even though, the decay of persistence probability is disconnected with the growth of domain length,
its behavior is strongly connected with the initial correlation length. 
It has been shown that the crossover length
to the second step of decays diverges as the equilibrium 
correlation length in both the dimensions. 
This leads to the question of difference in the fractal
dimensions in the pre and post crossover regimes which needs to be looked at carefully.
\par
We have not observed any initial configuration dependence of the growth of the 
average domain size. This is consistent with a previous study \cite{leticia} but more explicitly demonstrated here.
Essentially, even the transients are not affected due to change in initial temperature.
However, stronger finite-size effects are detected for lower values of $T_i$.
For domain growth, a striking observation is that the early 
time exponent is much higher than the asymptotic value, despite $T_f$ being zero.
This is at variance with the conserved order parameter dynamics.
These are all
interesting new results, requiring appropriate theoretical attention.
\par
In future we will focus on persistence for the conserved order parameter 
dynamics. For the conserved dynamics, initial temperature dependence of aging and 
domain growth are also important open problems.

\section*{Acknowledgement}\label{ack}
The authors are thankful to the Department of Science and Technology, India, for financial support via 
grant number SR/S2/RJN-13/2009.

~${*}$ das@jncasr.ac.in

\end{document}